\documentclass[10pt,conference,a4paper]{IEEEtran}

\IEEEoverridecommandlockouts

\usepackage{cite}
\usepackage{amsmath,amssymb,amsfonts}
\usepackage{algorithmic}
\usepackage{graphicx}
\usepackage{textcomp}
\usepackage{xcolor}
\usepackage{mathtools}

\usepackage{lipsum}
\usepackage[mathscr]{euscript}
\usepackage{algorithm}
\usepackage{multicol}
\usepackage{amsthm}

\theoremstyle{remark}
\newtheorem{theorem}{Theorem}
\newtheorem{proposition}[theorem]{Proposition}
\newtheorem{definition}[theorem]{Definition}
\newtheorem{remark}{Remark}
\newtheorem{assum}{Assumption}

\renewcommand{\Pr}{\mathbb{P}}

\allowdisplaybreaks

\renewcommand\IEEEkeywordsname{Keywords}

\begin{document}

	\title{Stochastic Model Predictive Control with Dynamic Chance Constraints\thanks{This work is supported by the Dutch NWO Veni project CODEC (project number 18244).}}
	
	\author{\IEEEauthorblockN{1\textsuperscript{st} M.H.W. Engelaar}
		\IEEEauthorblockA{\textit{Department of Electrical Engineering} \\
			\textit{Eindhoven University of Technology}\\
			Eindhoven, The Netherlands \\
			m.h.w.engelaar@tue.nl}
		\and
		\IEEEauthorblockN{2\textsuperscript{nd} S. Haesaert}
		\IEEEauthorblockA{\textit{Department of Electrical Engineering} \\
			\textit{Eindhoven University of Technology}\\
			Eindhoven, The Netherlands \\
			s.haesaert@tue.nl}
		\and
		\IEEEauthorblockN{3\textsuperscript{rd} M. Lazar}
		\IEEEauthorblockA{\textit{Department of Electrical Engineering} \\
			\textit{Eindhoven University of Technology}\\
			Eindhoven, The Netherlands \\
			m.lazar@tue.nl}
	}
	
	\maketitle
	
	
	\begin{abstract}
		
		This work introduces a stochastic model predictive control scheme for dynamic chance constraints. We consider linear discrete-time systems affected by unbounded additive stochastic disturbance. To synthesize an optimal controller, we solve two subsequent stochastic optimization problems. The first problem concerns finding the maximal feasible probabilities of the dynamic chance constraints. After obtaining the probabilities, the second problem concerns finding an optimal controller using stochastic model predictive control. We solve both stochastic optimization problems by reformulating them into deterministic ones using probabilistic reachable tubes and constraint tightening. We prove that the developed algorithm is recursively feasible and yields closed-loop satisfaction of the dynamic chance constraints. In addition, we will introduce a novel implementation using zonotopes to describe the tightening analytically. Finally, we will end with an example illustrating the method's benefits.
		
	\end{abstract}
	
	
	
	
	\section{Introduction}
	
	Stochastic model predictive control (SMPC) represents an effective control technique for the reliable handling of (chance) constraints in the presence of (unbounded) stochastic disturbances \cite{Mesb2016}. It has found applications in many areas, including vehicle path planning, building climate control, and power generation and distribution \cite{Mesb2016}. Typically stochastic model predictive control can be divided into two classes: \emph{randomized} methods and \emph{analytic approximation} methods \cite{Fari2016}. The former relies on generating sufficient disturbance realizations, while the latter reformulates the stochastic optimization problem into a deterministic optimization problem. 
	
	Regarding analytic approximation methods, existing techniques often consider \emph{static chance constraints}, e.g. \cite{Hewi2018,Kouv2016, Lore2016}. A disadvantage of these static chance constraints is that infeasibility at initialization often needs to be resolved by lowering the probability of the chance constraints over the entire horizon. To improve upon this, our paper will focus on \emph{dynamic chance constraints}. Moreover, we consider an optimization scheme that allows for relaxing the probability at specific instances to ensure feasibility at initialization. This allows for targeted chance constraint relaxation without requiring a worst-case relaxation over the entire horizon.
	
	Ensuring feasibility at initialization gives rise to several theoretical problems that must be addressed. First and foremost, contrary to existing work \cite{Hewi2018, Hewi2020, Kohl2022, Kouv2016, Schl2022}, we will have to solve two subsequent stochastic optimization problems. The first optimization problem concerns finding the maximal feasible probabilities on the dynamic chance constraints. After obtaining the probabilities, the second optimization problem concerns finding the optimal controller by solving an SMPC optimization problem. Hereby, we must develop a suitable terminal set that ensures recursive feasibility and closed-loop chance constraint satisfaction.
	
	To solve both stochastic optimization problems, we are specifically interested in utilizing probabilistic reachable sets (PRS), i.e., sets that satisfy the invariance property up to a given probability. Existing work regarding PRS includes the work done by \cite{Hewi2018, Hewi2020, Kohl2022, Schl2022}. These papers utilize sequences of \emph{static PRS} to reformulate static chance constraints on the system dynamics offline into static deterministic constraints on the nominal dynamics. These deterministic constraints are obtained by tightening the chance constraints utilizing PRS obtained from the probabilities on the chance constraints and the error dynamics. 
	Similarly, in this paper, we will utilize sequences of \emph{dynamic PRS}, called probabilistic reachable tubes (PRT), to obtain deterministic reformulations of both stochastic optimization problems, the former becoming a linear program and the latter becoming a tube-based MPC optimization problem. This will be our second contribution.
	
	
	As a final contribution, to compute the tightened constraints online, we will also develop a method based on properties of Minkowski set algebra to formulate tightened constraints analytically. Moreover, we provide a solution that circumvents using an ellipsoidal representation of the PRS, as these ellipsoidal representations are not tractable for Minkowski set difference operations. More specifically, we will over-approximate the ellipsoidal reachable sets using zonotopes, simplifying much of the tightening procedure at the cost of introducing some conservatism. Nevertheless, conservatism can be reduced by increasing the complexity of the zonotopes.
	
	
	In Section \ref{Sec:ProbSet}, we first introduce the problem setup of the paper. Next, in Section \ref{Sec:PRS}, we will define probabilistic reachable tubes, reformulate both stochastic optimization problems, show recursive feasibility and prove chance constraint satisfaction. Afterwards, in Section \ref{Sec:Impl}, we will discuss how to perform the tightening analytically utilizing zonotopes. In Section \ref{Sec:CaseStudy}, we will consider an example to illustrate the benefits of our method.

	\noindent \textbf{Notation:}
	The probability of $x\in A$, the expected value of random variable $x$ and the variance of random variable $x$ are written as $\Pr(x \in A), \mathbb{E}(x)$ and $\text{var}(x)$, respectively. The weighted 2-norm of a vector $x \in \mathbb{R}^n$ is denoted by $||x||_P^2=x^TPx$ for strictly positive definite matrices $P \in \mathbb{R}^{n \times n}$. The Pontryagin/Minkowski set difference of $A,B \subseteq \mathbb{R}^n$ is given by $A \ominus B := \{a \mid a+b \in A \ \forall b \in B\}$.
	
	
	\section{Problem Setup} \label{Sec:ProbSet}
	
	\noindent \textbf{Stochastic Linear Systems.}
	We consider a linear time-invariant (LTI) system with additive noise, given by
	\begin{equation} \label{Sys}
		x(k+1)=Ax(k)+Bu(k)+w(k),
	\end{equation}
	where $x \in \mathbb{X} \subseteq \mathbb{R}^n$ is the state, $x(0) \in \mathbb{X}$ is the initial state, $u \in \mathbb{U} \subseteq \mathbb{R}^m$ is the input and $w \in \mathbb{R}^n$ is an independent, identically distributed noise disturbance with distribution $w(k) \sim \mathcal{Q}^w$, which can have infinite support. We will assume that the disturbance $w(k) \sim \mathcal{Q}_w$ has at least known mean and variance and that the disturbance is central convex unimodal\footnote{$\mathcal{Q}_w$ is in the closed convex hull of all uniform distributions on symmetric compact convex bodies in $\mathbb{R}^n$ (c.f. \cite[Def. 3.1]{Dhar1976}).}. We say that a controller $\boldsymbol f$ is a sequence of policies $\boldsymbol f:= \{f_0, f_1, \dots\}$, such that $f_k$ maps states to inputs $f_k: \mathbb{X} \to \mathbb{U}$ for which the chosen control inputs for system \eqref{Sys} are given by $u(k)=\boldsymbol f(x(k))= f_k(x(k))$.
	\smallskip
	
	
	
	
	
	
	\noindent \textbf{Safety \& Performance.}	
	In this paper, we consider synthesis of a controller $\boldsymbol f$ for system \eqref{Sys} with safety and performance specifications. We will consider safety and performance specifications based on the probability that both state and input of \eqref{Sys} will remain within a specific safety set at each time step. Additionally, we will consider dynamic probabilities, which will result in dynamic chance constraints. The dynamic chance constraints are represented by
	\begin{subequations} \label{ChanCons}
		\begin{align}
			\mathbb{P}(x(k) \in \mathcal{X} {\mid} x(0)) &\geq \min(\bar p_x,p_x(k)), \\
			\mathbb{P}(u(k) \in \mathcal{U} {\mid}x(0)) &\geq \min(\bar p_u,p_u(k)),
		\end{align}
	\end{subequations}
	where $\mathcal{X}$ and $\mathcal{U}$ are convex sets containing the origin in their interior. Here $\bar p_x$ and $\bar p_u$ represent the \emph{target lower bounds}, i.e., the probability targets, and $p_x(k)$ and $p_u(k)$ represent the \emph{relaxed lower bounds} at time $k$. The relaxed lower bounds are defined as minimal or least costly relaxations of the target lower bounds, necessary for the existence of a controller $\boldsymbol f$. Finally, the constraints are defined with respect to the initial state, i.e., conditioned based on the initial state $x(0)$.

	To measure safety, we will consider a cost function that penalizes deviations of the relaxed lower bounds away from the target lower bounds, given by
	\begin{align} \label{CostFuncJp}
		J_p(\boldsymbol{p_x},\boldsymbol{p_u})= &\textstyle \sum_{k=0}^\infty \left(|\bar p_x-p_x(k)|
		+ |\bar p_u-p_u(k)| \right).
	\end{align}
	where $\boldsymbol{p_x}=\{ p_x(0), p_x(1), ...\}$ and $\boldsymbol{p_u}=\{ p_u(0), p_u(1), ...\}$. To measure the performance of a controller, we will penalize the distance between the state and the input with regard to the origin. Since system \eqref{Sys} is stochastic, similar to \cite{Kouv2016}, we consider the following cost function
	\begin{equation} \label{CostFuncJf}
		J_f(\boldsymbol x,\boldsymbol u)=\textstyle \sum_{k=0}^\infty \mathbb{E}_0\left(||x(k)||^2_{Q}+||u(k)||^2_{R}-l_{ss}\right),
	\end{equation}
	where $\boldsymbol x = \{x(0), x(1), \dots\}$ and $\boldsymbol u=\{u(0),u(1),\dots \}$ are sequences, $Q$ and $R$ are strictly positive definite matrices, $\mathbb{E}_0$ is the expected value conditioned on $x(0)$, and $l_{ss}$ is the expected infinite steady-state cost subtracted at each stage to ensure that the sum is finite.
	\smallskip
	
	
	
\begin{figure}[htp]
		\includegraphics[width=\columnwidth]{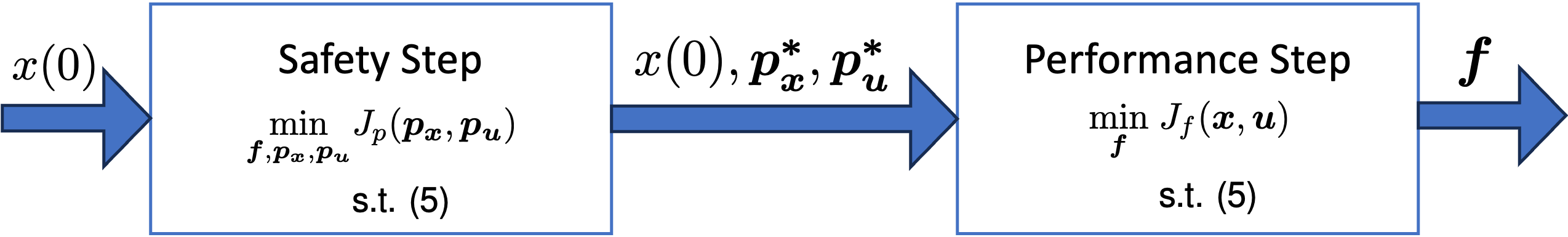}
		\centering
		\caption{Illustration of the two subsequent stochastic optimization problems.}
		\label{Fig1}
	\end{figure}

	\noindent \textbf{Problem Formulation.}	
	In this paper, we consider the problem of synthesizing a controller $\boldsymbol f$ that maximizes both safety and performance, i.e., given system \eqref{Sys} and constraints \eqref{ChanCons}, minimize cost functions \eqref{CostFuncJp} and \eqref{CostFuncJf}. We will reformulate this problem statement into two subsequent optimization problems, see also Figure \ref{Fig1}. The \emph{safety step} optimizes the safety allocation by means of optimization problem $\min_{\boldsymbol f,\boldsymbol{p_x},\boldsymbol{p_u}} J_p(\boldsymbol{p_x},\boldsymbol{p_u})$ such that
	\begin{subequations} \label{StocOptProb}
		\begin{align}
			x(k+1)&= Ax(k)+Bu(k)+w(k),\\
			u(k)&= f_k(x(k)), \ w(k) \sim Q_w, \\
			\mathbb{P}(x(k) &\in \mathcal{X} {\mid} x(0)) \geq p_x(k), \label{StocOptProbSta} \\
			\mathbb{P}(u(k) &\in \mathcal{U} {\mid}x(0)) \geq p_u(k), \label{StocOptProbInp} \\
			 p_x(k) &\in [0,\bar p_x], \ p_u(k) \in [0,\bar p_u]. \label{StocOptProbProb}
		\end{align}
	\end{subequations}
	Hereby, the safety step maximizes safety while ensuring a controller $\boldsymbol f$ exists. Given $\boldsymbol{p_x^*}$ and $\boldsymbol{p_u^*}$, the optimal solutions to the safety step, the \emph{performance step} then finds the best performing controller $\boldsymbol f$ by means of optimization problem $\min_{\boldsymbol f} J_f(\boldsymbol x,\boldsymbol u)$ such that \eqref{StocOptProb} holds true, ignoring \eqref{StocOptProbProb}.

	
	\noindent \textbf{Approach.}
	To solve both stochastic optimization problems in a tractable way, we will build upon existing methods, such as the effective framework based on stochastic model predictive control (SMPC) with constraint tightening from \cite{Hewi2018,Hewi2020,Schl2022}. Therein, it was explained how to synthesize a stochastic model predictive controller $\boldsymbol{f}$ for the performance step, assuming that possible relaxation of the target lower bounds is ignored, i.e., by assuming that $p_x(k)=\bar p_x$ and $p_u(k)=\bar p_u$. This was accomplished using static probabilistic reachable sets to obtain a deterministic tube-based MPC reformulation of the SMPC optimization problem.
	
	In this work, we will expand upon this by considering dynamic probabilistic reachable sets, that is, probabilistic reachable sets that differ in time, called probabilistic reachable tubes. To obtain the initial tubes over an infinite horizon, we first solve for the safety step by reformulating the safety step as a deterministic linear program. Utilizing the initialized tubes, we solve for the performance step by reformulating the performance step as a deterministic tube-based MPC optimization problem. During the performance step, we will update the tubes at each prediction step as part of a receding horizon. 
	The main advantage of our method is that it allows for optimal safety and feasibility due to the tubes' dynamic nature while also having the ability to optimize performance.

	\section{Probabilistic Reachable Tubes \& Deterministic tube-based MPC Reformulation} \label{Sec:PRS}
	
	First, let us define probabilistic reachable tubes together with their ellipsoidal explicit representation to afterwards explain how deterministic reformations for both the safety and performance steps can be obtained similar to \cite{Hewi2018,Hewi2020,Schl2022}. At the end of this section, we will prove that the closed-loop system satisfies chance constraints \eqref{ChanCons} and that the deterministic reformulations together are recursively feasible.
	
	
	\subsection{Probabilistic Reachable Tubes}
	
	Consider an autonomous stochastic linear system given by
	\begin{align} \label{SysProbReachSet}
		x(k+1)= A_K x(k) +w(k),
	\end{align}
	where $x \in \mathbb{R}^n$ is the state, $A_K=A+BK$, $w(k) \sim \mathcal{Q}^w$ is the disturbance, and $K$ is a feedback controller meant to stabilize the system, i.e., $A_K$ has eigenvalues strictly inside the unit circle. According to \cite{Hewi2018}, the definition of the probabilistic reachable set for a \emph{static} probability level is given as follows.
	\begin{definition}[Probabilistic reachable sets] \label{Def:PRS}
		A set $\mathcal{R}$ is said to be a probabilistic reachable set (PRS) of probability level $p \in [0,1]$ for system \eqref{SysProbReachSet} if
		\begin{equation} \label{PRSReq}
			x(0) = 0 \implies \Pr(x(k) \in \mathcal{R}) \geq p, \ \forall k \geq 0.
		\end{equation}
	\end{definition}
	\smallskip
	
	Assume that we have \emph{dynamic} probability levels. We will define the probabilistic reachable tubes as follows.
	\begin{definition}[Probabilistic reachable tubes] \label{Def:PRT}
		A probabilistic reachable tube (PRT) of dynamic probability levels $\boldsymbol p=\{p(0),p(1),\dots\}$ for system \eqref{SysProbReachSet}, denoted by $\mathcal{R}^{\boldsymbol p}$, is a sequence of PRS, where the $k^{\text{th}}$ element is a PRS of probability level $p(k)$ for system \eqref{SysProbReachSet}.
	\end{definition}
	
	To obtain an explicit form for any PRS, multiple approaches exist, see \cite{Hewi2018,Hewi2020}. In this paper, we will consider the popular ellipsoidal explicit representation \cite{Hewi2018}. For simplicity, we will make the following assumption.
	\begin{assum} \label{Ass:Mean}
		Disturbance $\mathcal{Q}_w$ has zero mean and has strictly positive definite variance.
	\end{assum}
	\noindent The above assumption is not necessary to obtain an ellipsoidal explicit representation but will simplify computation. 
	
	The ellipsoidal explicit representation is obtained from the multivariable Chebyshev inequality, details are given in \cite{Hewi2018}. Under Assumption \ref{Ass:Mean}, the ellipsoidal explicit representation of a PRS $\mathcal{R}$ of probability level $p$ for system \eqref{SysProbReachSet} is given by
	\begin{equation}\label{EllipNot}
		\mathcal{R}= \{x \in \mathbb{R}^{n} \mid x^T \Sigma_{\infty}^{-1}x \leq \tilde p\}=\mathcal{E}(\tilde p \Sigma_{\infty},0),
	\end{equation}
	where $\mathcal{E}(E,\bar e):=\{x \in \mathbb{R}^n {\mid} (x-\bar e)^TE^{-1}(x-\bar e) \leq 1\}$, with $E \in \mathbb{R}^{n \times n}$ being strictly positive definite and $\bar e \in \mathbb{R}^n$ being the center of the ellipsoid; $\Sigma_{\infty}$ solves the Lyapunov equation \begin{equation}
		A_K\Sigma_{\infty}A_K^T-\Sigma_{\infty}+\text{var}(\mathcal{Q}_w)=0;
	\end{equation}
	and $\tilde p=\frac{n}{1-p}$, or, in the case of Gaussian disturbance, $\tilde p=\chi^2_{n}(p)$, the inverse cumulative distribution function of the chi-squared distribution with $n$ degrees of freedom. Hereby, the latter $\tilde p$ will yield a less conservative PRS in case of a Gaussian disturbance $w$. 
	
	Consider the state space transformation
	\begin{equation} \label{SysExt}
		x_K(k)=Kx(k),
	\end{equation}
	where $x(k)$ is the state vector in \eqref{SysProbReachSet}. Under Assumption \ref{Ass:Mean}, we have that $\mathcal{R}_K = \mathcal{E}(\tilde p K\Sigma_{\infty}K^T,0)$ is a PRS of probability level $p$ for the trajectory given by \eqref{SysExt}, i.e., $\mathcal{R}_K$ is such that
	\begin{equation}
		x_K(0) = 0 \implies \Pr(x_K(k) \in \mathcal{R}_K) \geq p, \ \forall k \geq 0.
	\end{equation}
	Here, $\Sigma_{\infty}$ and $\tilde p$ are obtained from \eqref{EllipNot}. Details can be found in \cite{Hewi2020,Schl2022}. We will make the following assumption.
	\begin{assum} \label{Ass:PRS}
		All PRS $\mathcal{R}$ and $\mathcal{R}_K$ will be considered exclusively in ellipsoidal explicit representation.
	\end{assum}
	
	\begin{remark} \label{Rem:PRT}
		With the ellipsoidal explicit representation, any probabilistic reachable tube will be a tube of ellipsoids centred around the origin. Each element of the tube will be a probabilistic reachable set of similar shape and orientation but of different sizes. 
		Any cross-section of the tube will give dynamic bounds on the corresponding system dynamics obtained from the dynamic probability levels. This is in contrast to methods such as \cite{Hewi2018,Hewi2020,Schl2022}, which obtain tubes for which the cross-section has static bounds on the system dynamics due to a constant probability level.
	\end{remark}
	
	
	\subsection{Deterministic Tube-Based MPC Reformulation}
	
	Following the approach in \cite{Hewi2018,Hewi2020,Schl2022}, the deterministic reformulations for both the safety and performance steps can be obtained as follows. 
	First, decompose the dynamics into a nominal and an error part:
	\begin{subequations}
		\begin{align}
			x(k) &= z(k)+ e(k), \label{Decomp:State}\\
			z(k+1) & = Az(k)+Bv(k),\\
			e(k+1) &= A_Ke(k)+w(k),\\
			u(k) &= v(k) +Ke(k) = v(k) +e_u(k).\label{Decomp:Inp}
		\end{align}
	\end{subequations}
	Here $K$ is the auxiliary stabilizing feedback controller meant to keep the error $e$ small. Next, determine the PRS $\mathcal{R}_x^k$ of probability level $p_x(k)$ for error $e$. Using decomposition \eqref{Decomp:State}, the chance constraint \eqref{StocOptProbSta} can now be reformulated as
	\begin{align} \label{RefChanCon}
		p_x(k) &\leq \Pr(z(k)+\mathcal{R}_x^k \subseteq \mathcal{X}\wedge e(k) \in \mathcal{R}_x^k)\\
		&\leq \Pr(z(k)+e(k) \in \mathcal{X})=\Pr(x(k) \in \mathcal{X}), \nonumber
	\end{align}
	i.e., a deterministic constraint for the nominal state $z$ together with a PRS $\mathcal{R}_x^k$ of probability level $p_x(k)$ for error $e$. Next, determine the probabilistic reachable set $\mathcal{R}_u^k$ of probability level $p_u(k)$ for input error $e_u$. Similarly, this allows for a reformulation of chance constraint \eqref{StocOptProbInp}.
	Together this yields the deterministic dynamic tightened constraints
	\begin{subequations} \label{DynDetCon}
		\begin{align}
			z(k) &\in \mathcal{Z}(k):=\mathcal{X}\ominus \mathcal{R}_x^k,\\
			v(k) &\in \mathcal{V}(k):=\mathcal{U}\ominus \mathcal{R}_u^k.
		\end{align}
	\end{subequations}
	
	Notice that the sequence $\mathcal{R}^{\boldsymbol{p_x}}=\{ \mathcal{R}_x^0, \mathcal{R}_x^1, \dots\}$ is a PRT with ellipsoidal elements. More specifically, Assumption \ref{Ass:PRS} and Remark \ref{Rem:PRT} imply that this is a sequence of ellipsoids centred around the origin and whereby each element of the sequence only differs in size. Accordingly, an alternative notation for $\mathcal{R}_x^k$ is given by $\alpha(k)\mathcal{R}_x^{\bar p}$, where $\alpha(k) \in [0,1]$ and $\mathcal{R}_x^{\bar p}$ is a PRS of probability level $\bar p_x$ for error $e$. To derive $\alpha(k)$, first notice that
	\begin{align*}
		\alpha(k)\mathcal{R}_x^{\bar p} &= \{x \in \mathbb{R}^{n} \mid x^T \Sigma_{\infty}^{-1}x \leq \alpha(k)^2\tilde p_x\},\\
		\mathcal{R}_x^k &= \{x \in \mathbb{R}^{n} \mid x^T \Sigma_{\infty}^{-1}x \leq \tilde p_x^k\}
	\end{align*}
	where $\tilde p_x$ is obtained from the ellipsoidal explicit representation of $\mathcal{R}_x^{\bar p}$, and $\tilde p_x^k$ is obtained from the ellipsoidal explicit representation of $\mathcal{R}_x^k$. Since we want that $\mathcal{R}_x^k=\alpha(k)\mathcal{R}_x^{\bar p} $, we derive that $\alpha=\sqrt{\frac{\tilde p_x^k}{\tilde p_x}}$. The same can be derived with regards to sequence $\mathcal{R}^{\boldsymbol{p_u}}=\{\mathcal{R}_u^0, \mathcal{R}_u^1, \dots \}$.
	
	The result is that we can reformulate the deterministic dynamic tightened constraints \eqref{DynDetCon} as
	\begin{subequations} \label{DynDetConPlus}
		\begin{align}
			\mathcal{Z}(k) &= \mathcal{X}\ominus(1-\alpha(k))\mathcal{R}_x^{\bar p},\\
			\mathcal{V}(k) &= \mathcal{U}\ominus (1-\beta(k))\mathcal{R}_u^{\bar p},
		\end{align}
	\end{subequations}
	whereby $\alpha(k)=\beta(k)=0$ will give the fully tightened constraints $\mathcal{Z}=\mathcal{X}\ominus \mathcal{R}_x^{\bar p}$ and $\mathcal{V}=\mathcal{U}\ominus \mathcal{R}_u^{\bar p}$, respectively; $\alpha(k)=\beta(k)=1$ will give the original constraints $\mathcal{X}$ and $\mathcal{U}$, respectively; and $0<\alpha(k)<1$ and $0<\beta(k)<1$ will give $\mathcal{Z} \subset \mathcal{Z}(k) \subset \mathcal{X}$ and $\mathcal{V} \subset \mathcal{V}(k) \subset \mathcal{U}$, respectively. All-in-all, optimizing over $\alpha$ and $\beta$ allows us to optimize over the relaxed lower bounds, whereby minimizing over $\alpha$ and $\beta$ will maximize the relaxed lower bounds and thus maximize safety.
	
	\begin{remark} \label{Rem:tightening}
		If $\alpha(k)$ and $\beta(k)$ are given, the corresponding PRS $\mathcal{R}_x^k$ and $\mathcal{R}_u^k$ can be determined and vice versa. Hence, throughout this paper, we can consider dynamic tightenings \eqref{DynDetCon} and \eqref{DynDetConPlus} as equivalent.
	\end{remark}

	To ensure that the infinite horizon optimization problems for the safety and performance steps can be solved numerically, we introduce a finite horizon $N$ and a terminal set $\mathcal{Z}_F$. We assume that the terminal set $\mathcal{Z}_F$, with respect to input $v(k)=Kz(k)$, satisfies $(A+BK)\mathcal{Z}_F \subseteq \mathcal{Z}_F \subseteq \mathcal{Z}$ and $K\mathcal{Z}_F \subseteq \mathcal{V}$. $\mathcal{Z}_F$ can be obtained from \cite[Theorem 2.3]{Kouv2016}. Notice that the terminal set is obtained based on maximal tightening. This will ensure that cost function \eqref{CostFuncJp} will be equal to the sum of its first $N$ terms. Additionally, as will be shown in the next subsection, this choice of terminal set will ensure recursive feasibility.
	
	\begin{figure}[htp]
		\vspace*{-0.4cm}\includegraphics[width=\columnwidth]{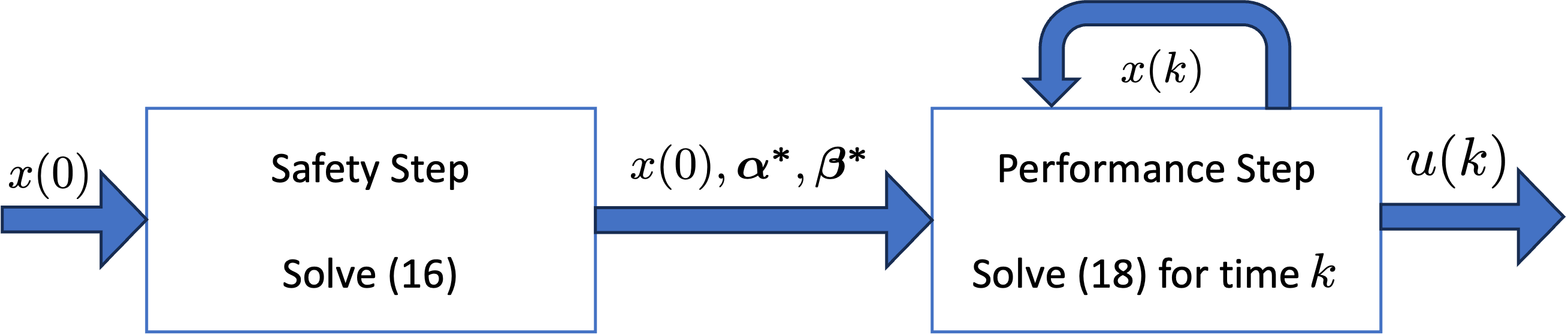}
		\centering
		\caption{Illustration of the safety step and the subsequent performance step. The former is solved only once, while the latter is solved for each time $k$.}
		\label{Fig2}
	\end{figure}
	
	All of the above now culminates into deterministic reformulations for both the safety and performance steps, see also Figure \ref{Fig2}. Let $\bar v=\{v(0), \dots, v(N-1)\}$, $\boldsymbol \alpha=\{\alpha(0), \dots, \alpha(N-1)\}$ and $\boldsymbol \beta=\{\beta(0), \dots,  \beta(N-1)\}$. The safety step is reformulated into a linear program given by  
	\begin{subequations} \label{LinProg}
		\begin{align}
			\min_{\bar v,\boldsymbol \alpha,\boldsymbol \beta} &\bar J_p(\boldsymbol \alpha,\boldsymbol \beta)\\
			\text{s.t.} \quad z(k+1)&= Az(k)+Bv(k), \label{LinProg:ConDyn}\\
			z(k) & \in \mathcal{X}\ominus (1-\alpha(k))\mathcal{R}_x^{\bar p}, \label{LinProg:ConState}\\
			v(k) & \in \mathcal{U}\ominus (1-\beta(k))\mathcal{R}_u^{\bar p},\label{LinProg:ConInp}\\
			z(N) & \in \mathcal{Z}_F, \ z(0) =x(0),\\
			\alpha(k) &\in [0,1], \ \beta(k)  \in [0,1], \\
			\forall k &\in \{0, \ldots, N-1\}.
		\end{align}
	\end{subequations}
	Here, the cost function is given by
	\begin{equation} \label{DetCostFuncJp}
		\bar J_p(\boldsymbol \alpha, \boldsymbol \beta)= \textstyle \sum_{k=0}^{N-1} \left[\alpha(k)+\beta(k)\right].
	\end{equation}
	
	Let $\boldsymbol{\alpha^*}=\{\alpha^*(0),\alpha^*(1),...\}$ and $\boldsymbol{\beta^*}=\{\beta^*(0), \beta^*(1),...\}$ be infinite sequences for which the first $N$ elements are given as the optimal solution to optimization problem \eqref{LinProg} and the remaining elements are equal to zero. Let additionally, for any vector or scalar of variables/signals $s(k)$, $s_i(k)$ denote the corresponding predicted value of $s(k+i)$ and $\bar s(k):=\{s_0(k), \ldots, s_{N-1}(k)\}$, all predicted based on measurements available at time $k$. Utilizing $\boldsymbol{\alpha^*}$ and $\boldsymbol{\beta^*}$, the performance step can now be reformulated into a deterministic tube-based MPC optimization problem given by
	\begin{subequations} \label{OptDetRef}
		\begin{align}
			\min_{\bar v(k), \xi(k)} &\bar J_f(\bar z(k),\bar v(k))+l(\xi(k))\\
			\text{s.t.} \quad z_{i+1}(k)&= Az_i(k)+Bv_i(k), \label{OptDetRef:ConDyn}\\
			z_i(k) & \in \mathcal{X}\ominus (1-\alpha^*(k+i))\mathcal{R}_x^{\bar p}, \label{OptDetRef:ConState}\\
			v_i(k) & \in \mathcal{U}\ominus (1-\beta^*(k+i))\mathcal{R}_u^{\bar p},\label{OptDetRef:ConInp}\\
			\forall i &\in \{0, \ldots, N-1\}, \label{OptDetRef:Iter}\\
			z_0(k) & =(1-\xi(k))x(k)-\xi(k)z_1(k-1), \label{OptDetRef:IniNomSta}\\
			z_{N}(k) & \in \mathcal{Z}_F, \ \xi(k) \in \{0,1\} \label{OptDetRef:TermXi}.
		\end{align}
	\end{subequations}
	Here, the cost function is given by
	\begin{equation} \label{DetCostFuncJf}
		\bar J_f(\bar z(k),\bar v(k))=\textstyle \sum_{i=0}^{N-1}||v_i(k)-Kz_i(k)||^2_{S}, 
	\end{equation}
	where $S=R+B^TPB$, $P$ is a stabilizing solution to the discrete algebraic Riccati equation and $K$ the stabilizing state-feedback gain, see also \cite{Ione1993}, given by
	\begin{subequations}\label{DARE}
		\begin{align}
			K&=-(R+B^TPB)^{-1}B^TPA,\\
			P&=A^TPA+A^TPBK+Q.
		\end{align}
	\end{subequations}
	
	To clarify, both optimization problems assume $K$ to be given by \eqref{DARE} and have the same terminal set $\mathcal{Z}_F$. Further, \eqref{DetCostFuncJp} is obtained from \eqref{CostFuncJp} and the assumption that after $N$ steps, the relaxed lower bounds become equal to the target lower bounds, i.e., maximal safety is assumed after $N$ steps. Next, \eqref{DetCostFuncJf} and \eqref{DARE} are obtained from substituting equations \eqref{Decomp:State} and \eqref{Decomp:Inp} into equation \eqref{CostFuncJf} and applying \cite[Corollary 6.1]{Kouv2016} after which constants are ignored within the cost function. Our choice of initial state $z_0(k)$ comes from the effective idea put forward in \cite{Hewi2018}, and it is assumed that $z_1(-1)=x(0)$ and $l(\xi(k))$ is either a linear or quadratic function that penalizes high values of $\xi(k)$.
	
	
	
	
	\begin{remark}
		In reformulation \eqref{OptDetRef}, the probabilistic reachable tubes have dynamic cross-sections. For each time step $k\leq N$, the cross-section of the probabilistic reachable tube might be updated. This is in contrast to existing SMPC methods such as \cite{Kouv2016, Hewi2018,Hewi2020,Schl2022}, for which the tube is time-invariant.
	\end{remark}
	
	
	\subsection{Recursive Feasibility \& Chance Constraint Satisfaction}
	
	Consider the following theorem regarding recursive feasibility of deterministic reformulations \eqref{LinProg} and \eqref{OptDetRef}.
	\begin{theorem}[Recursive Feasibility] \label{Thm:RecuFeas}
		Given a feasible solution of \eqref{LinProg} exists, the tube-based MPC optimization problem in \eqref{OptDetRef} is recursively feasible.
	\end{theorem}
	\begin{proof}
		First notice that any solution $\bar v$ of \eqref{LinProg} is also a solution of \eqref{OptDetRef} at time $k=0$. We will show that if a solution of \eqref{OptDetRef} exist at time $k$ this implies a solution exists at time $k+1$, thereby proving recursive feasibility by induction. Let $\bar v(k)=\{v_0(k), ..., v_{N-1}(k)\}$ be any solution to \eqref{OptDetRef} at time $k$ and let $\bar z(k)=\{z_0(k), ..., z_N(k)\}$ be the corresponding nominal states. Take $\bar v(k+1) =\{v_1(k), ..., v_{N-1}(k),Kz_N(k)\}$ and $\xi(k+1)=1$. According to \eqref{OptDetRef:ConDyn} and \eqref{OptDetRef:IniNomSta}, we have that $\bar z(k+1)=\{z_1(k), ..., z_N(k),A_Kz_N(k)\}$. Remember that $A_K \mathcal{Z}_F \subseteq \mathcal{Z}_F$, hence $z_N(k+1)=A_Kz_N(k) \in \mathcal{Z}_F$. All-in-all, we have proven satisfaction of constraints \eqref{OptDetRef:ConDyn}, \eqref{OptDetRef:Iter}, \eqref{OptDetRef:IniNomSta} and \eqref{OptDetRef:TermXi} at time $k+1$. Consider, next constraint \eqref{OptDetRef:ConState}. For $i \in \{0,\dots, N-2\}$ satisfaction of constraint \eqref{OptDetRef:ConState} follows trivially. For $i=N-1$, first remember that $\alpha^*(k+N)=0$ and $z_{N-1}(k+1)=z_N(k) \in \mathcal{Z}_F \subseteq \mathcal{Z}=\mathcal{X}\ominus \mathcal{R}_x^{\bar p}$. Satisfaction of constraint \eqref{OptDetRef:ConState} now follows trivially. Consider last constraint \eqref{OptDetRef:ConInp}. For $i \in \{0,\dots, N-2\}$ satisfaction of constraint \eqref{OptDetRef:ConInp} follows trivially. For $i=N-1$, first remember that $\beta^*(k+N)=0$, $K\mathcal{Z}_F \subseteq \mathcal{V}$ and $v_{N-1}(k+1)=Kz_N(k) \in \mathcal{V}=\mathcal{U}\ominus \mathcal{R}_u^{\bar p}$. Satisfaction of constraint \eqref{OptDetRef:ConInp} now follows trivially. Hence, $\bar v(k+1)$ is a (non-optimal) solution of \eqref{OptDetRef} at time $k+1$, implying also an optimal solution exists, thereby finishing the proof.
	\end{proof}
	
	To prove chance constraint satisfaction, first, consider the following proposition obtained from \cite[Theorem 3]{Hewi2018}.
	\begin{proposition} \label{Prop}
		Let $\mathcal{Q}_w$ be central convex unimodal, and let $\mathcal{R}$ be a PRS of probability level $p$ for error $e$. For system \eqref{Sys} under the control law \eqref{Decomp:Inp} resulting from \eqref{OptDetRef} with tightening \eqref{DynDetConPlus}, we have
		\begin{equation}
			\Pr(e_0(k)\in \mathcal{R}) \geq \Pr(e_k(0) \in \mathcal{R})
		\end{equation}
		for all $k \geq 0$, conditioned on $e(0)=e_0(0)=0$.
	\end{proposition}
	The proof of the above proposition follows directly from the proof of \cite[Theorem 3]{Hewi2018} and Remark \ref{Rem:tightening}.
	
	Consider now the following theorem regarding the satisfaction of the chance constraints \eqref{StocOptProbSta} and \eqref{StocOptProbInp} via the deterministic tube-based MPC reformulation \eqref{OptDetRef}.
	\begin{theorem}[Change Constraint Satisfaction]
		The system \eqref{Sys} under the control law \eqref{Decomp:Inp} resulting from \eqref{OptDetRef} with tightening \eqref{DynDetConPlus}, will satisfy chance constraints \eqref{StocOptProbSta} and \eqref{StocOptProbInp}.
	\end{theorem}
	\begin{proof}
		First notice that $z_0(0)=x(0)$ implies that $e(0)=0$. Let $\mathcal{R}_x^k$ be a PRS of probability level $p_x(k)$ for error $e$. According to Definition \ref{Def:PRS}, we have that $\Pr(e_i(0) \in \mathcal{R}_x^k)\geq p_x(k)$ for all $i,k \in \{0,1,\dots\}$. Utilizing Proposition \ref{Prop}, we have that $\Pr(e_0(k) \in \mathcal{R}_x^k)\geq p_x(k)$, where we took $i=k$. Next recall from Theorem \ref{Thm:RecuFeas} that $z(k)=z_0(k) \in \mathcal{X}\ominus \mathcal{R}_x^k$. Satisfaction of chance constraint \eqref{StocOptProbSta} now follows directly from \eqref{RefChanCon}. Satisfaction of chance constraint \eqref{StocOptProbInp} can be proven in a similar manner.
	\end{proof}
	
	
	
	\section{Implementation} \label{Sec:Impl}
	
	The main difficulty of implementing \eqref{LinProg} and \eqref{OptDetRef} is with regards to \eqref{LinProg:ConState}, \eqref{OptDetRef:ConState}, \eqref{LinProg:ConInp}, and \eqref{OptDetRef:ConInp}. Hence, in this section, we will explain how to obtain these analytically. We will achieve this by first taking a zonotopic over-approximation of the ellipsoidal explicit representation, to next rewrite the zonotope into vertex representation, after which the Pontryagin set difference can be described analytically. We will assume the following with regard to constraints $\mathcal{X}$ and $\mathcal{U}$ in \eqref{ChanCons}.
	\begin{assum}
		$\mathcal{X}$ and $\mathcal{U}$ have known half-space representation given by $\mathcal{X}=\{x \in \mathbb{R}^n \mid A_xx\leq b_x\}$ and $\mathcal{U}=\{u \in \mathbb{R}^m \mid A_uu\leq b_u\}$.
	\end{assum}
	
	First, consider a PRS $R$ given by the ellipsoidal explicit representation $\mathcal{E}(\tilde pE,0)$. A zonotopic over-approximation of $\mathcal{R}$ can be obtained from \cite[Theorem 4]{Gass2020}. 
	Next, we use \cite[Theorem 2]{Alth2015} to rewrite the zonotope into a half-space representation $\{x \in \mathbb{R}^n \mid A_zx\leq b_z\}$, where $A_z \in \mathbb{R}^{q \times n}$ and $b_z \in \mathbb{R}^q$. Finally, we can rewrite the zonotope into vertex representation $\{x \in \mathbb{R}^n \mid x=\sum_{i=1}^r a_iv_i \text{ s.t. } \forall i, \  a_i\geq 0, \ \sum_{i=1}^r a_i=1\}$ using the algorithm given by \cite{Avis1991}. The vertex representation with vertex set $\mathscr{V}=\{v_1,\ldots, v_r\}$ then allows us to use \cite[Theorem 2.1(xiii)]{Kolm1998}, to conclude that $$\mathcal{X}\ominus\alpha\mathcal{R}=\textstyle \bigcap_{v \in \mathscr{V}}\mathcal{X}-\alpha v,$$ where $\forall v \in \mathscr{V}$ we have that
	\begin{align*}
		\mathcal{X}-\alpha v:&=\{x-\alpha v \in \mathbb{R}^n \ \mid A_x x \leq b_x\}\\
		&= \{y \in \mathbb{R}^n \mid A_xy\leq b_x - \alpha A_x v\}.
	\end{align*}
	The same can be concluded with regards to $\mathcal{U}$. As a result, we can rewrite \eqref{LinProg:ConState}, \eqref{LinProg:ConInp}, \eqref{OptDetRef:ConState}, and \eqref{OptDetRef:ConInp} into a finite number of linear inequalities.
	
	\begin{remark}
		By over-approximating ellipsoids by zonotopes, conservatism will be introduced. Nevertheless, this can be mitigated by over-approximating tighter and tighter the ellipsoids.
	\end{remark}
	
	
	
	\section{Case Study} \label{Sec:CaseStudy}
	
	To illustrate our method, we will consider the benchmark case study in the SMPC literature of the DC-DC-converter regulation problem \cite{Cann2010,Schl2022}. Here, the linear dynamics are of the form \eqref{Sys} given by
	\begin{equation}
		A=\begin{bmatrix}
			1.000 & 0.0075\\ -0.143 & 0.996
		\end{bmatrix}, \quad 
		B=\begin{bmatrix}
			4.798\\ 0.115
		\end{bmatrix},
	\end{equation}
	where we assume that the disturbance is Gaussian with zero mean and variance $0.1I_2$. We assume that the performance cost function has weights $Q=\text{diag}[1, 10]$ and $R=10$ and the prediction horizon will be $N=15$. Finally, we consider a chance constraint on each element of the state given by 
	\begin{equation}
		\Pr(-2 \leq x^i(k)\leq 2)\geq 0.6, \ i \in \{1,2\},
	\end{equation}
	that is, the target lower bound is given by $\bar p_x=0.6$, each element of the state must be within the interval $[-2,2]$ and no constraints on the input.
	
	\begin{figure}[htp]
		\centering
		\vspace*{-0.2cm}\hspace*{-0.3cm}\includegraphics[trim={1.8cm 0cm 3cm 0.5cm},clip,width=1\columnwidth]{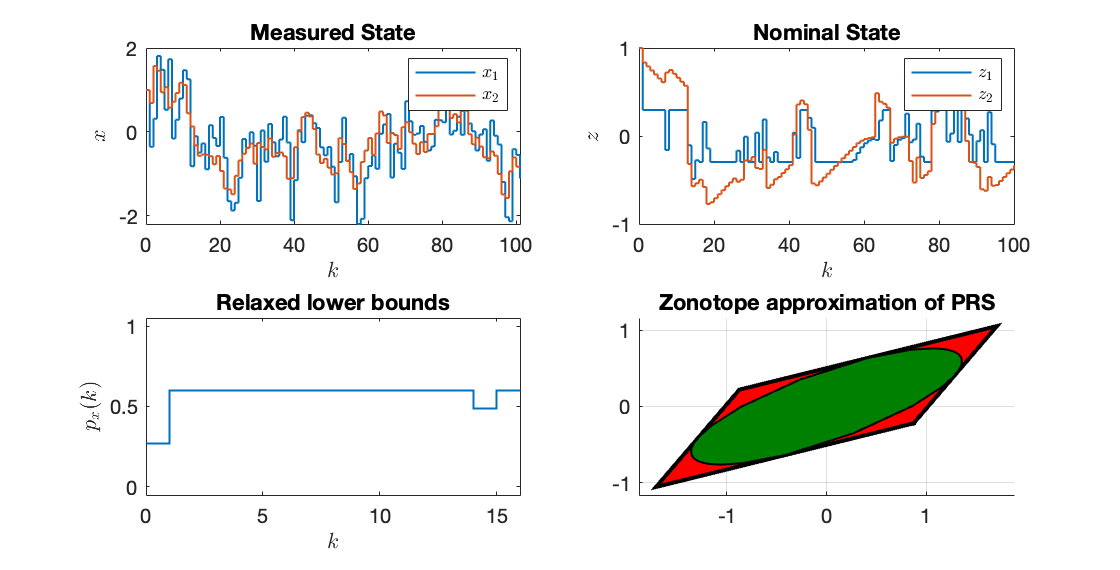}
		\vspace*{-.4cm}\caption{\textbf{Top-Left:} The measured state over a time horizon $[0,100]$. \textbf{Top-Right:} The nominal state over a time horizon $[0,100]$. \textbf{Lower-Left:} The relaxed lower bound calculated during the safety step. \textbf{Lower-Right:} The zonotope over-approximation of a probabilistic reachable set.}
		\label{Fig:Results}
	\end{figure}
	
	To showcase the benefits of our method, we consider an initial state for which the choice of static chance constraints will result in infeasibility, i.e. no controller can be synthesized without lowering the target lower bound. One such initial condition is given by $x(0)=[1,1]^T$. As explained in this paper, a safety allocation is first attempted by solving the safety step optimization problem \eqref{LinProg}. If this attempt is met positively, the controller can be determined by solving the performance step optimization problem \eqref{OptDetRef}. The results of both optimization problems can be found in Figure \ref{Fig1}, and the probabilistic reachable tube corresponding to the relaxed lower bounds can be observed in Figure \ref{Fig2}. 
		
		\begin{figure}[htp]
			\hspace*{-0.15cm}\includegraphics[trim={1.8cm 0cm 2.0cm 0},clip,width=\columnwidth]{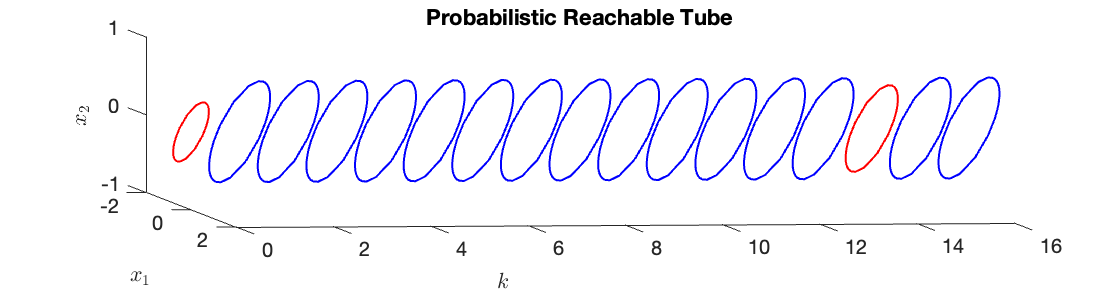}
			\centering
			\vspace*{-0.3cm}\caption{The probabilistic reachable tube corresponding to the relaxed lower bounds in Figure \ref{Fig1}. Notice that at time $k=0$ and $k=14$ the red ellipsoids are smaller, in correspondence to the relaxed lower bounds in Figure \ref{Fig1}.}
			\label{Fig:Results2} 
		\end{figure}
		
	From Figure \ref{Fig1}, it can be deduced that relaxation occurred at time $k=0$ and at time $k=14$. The former is due to our choice of the initial state, while the latter is due to enforcing the nominal state to be within the terminal set after $N=15$ time steps. That relaxation happens at both time instances can also be observed from the nominal trajectory $z_1$. Notice that at time $k=0$ and time $k=14$, the nominal state $z_1$ is respectively a maximum and a minimum, violating the boundary enforced by the target lower bound. All-in-all, it can be concluded that initial state $x(0)$ would lead to infeasibility should the chance constraints be static, as relaxation of the target lower bound would be necessary. More specifically, the target lower bound would have to be lowered to $\bar p_x=0.27$ for a solution to exist.

	\section{Conclusion \& Future Work} \label{Sec:Conc}
	
	In this work, we have introduced a stochastic model predictive control scheme for dynamic chance constraints. We considered two subsequent stochastic optimization problems, the first of which optimizes safety and the second of which optimizes performance. By utilizing probabilistic reachable tubes with dynamic cross-sections, we have tightened the dynamic chance constraint, thereby reformulating both stochastic optimization problems into, respectively, a linear program and a tube-based MPC optimization problem. We have shown that the deterministic reformulations are recursively feasible, and the closed-loop system satisfies the dynamic chance constraints. In addition, we have introduced a novel implementation using zonotopes to describe the tightening analytically. Finally, we gave an example to illustrate the method's benefits. In the future, we also want to prove stability and consider lexicographic formulations of the problem statement.
	
	
	\bibliographystyle{IEEEtranS}
	\bibliography{ArticleSMPCFV}
	
\end{document}